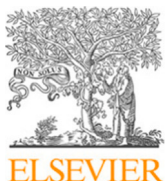
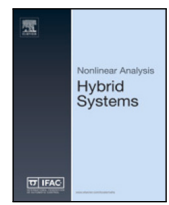

# Time- versus event-triggered consensus of a single-integrator multi-agent system

David Meister [a],[*], Frank Aurzada [b], Mikhail A. Lifshits [c], Frank Allgöwer [a]

[a] *University of Stuttgart, Institute for Systems Theory and Automatic Control, Germany*
[b] *Technical University of Darmstadt, Germany*
[c] *St. Petersburg State University, Russia*

A B S T R A C T

Event-triggered control has shown the potential for providing improved control performance at the same average sampling rate when compared to time-triggered control. While this observation motivates numerous event-triggered control schemes, proving it from a theoretical perspective has only been achieved for a limited number of settings. Inspired by existing performance analyses for the single-loop case, we provide a first fundamental performance comparison of time- and event-triggered control in a multi-agent consensus setting. For this purpose, we consider undirected connected network topologies without communication delays, a level-triggering rule for event-triggered control, and the long-term average of the quadratic deviation from consensus as a performance measure. The main finding of our analysis is that time-triggered control provably outperforms event-triggered control beyond a certain number of agents in our particular setting. We thereby provide an illustrative distributed problem setup in which event-triggered control results in a performance disadvantage when compared to time-triggered control in the case of large networks. Moreover, we derive the asymptotic order of the performance measure under both triggering schemes which gives more insights into the cost relationship for large numbers of agents. Thus, by presenting an analysis for a particular setup, this work points out that transferring an event-triggering scheme from the single-loop to the multi-agent setting can lead to a loss of the often presumed superiority of event-triggered control over time-triggered control. In particular, the design of performant decentralized event-triggering schemes can therefore pose additional challenges when compared to the analogue single-loop case.

## 1. Introduction

Event-triggered control (ETC) schemes have shown the potential to be more performant than time-triggered control (TTC) schemes when communication channels are loss- and delay-free, and average triggering rates are equal, as demonstrated in [1] for single-integrator systems. In ETC, the system only communicates when a triggering condition is met, whereas in TTC, it communicates at fixed time intervals. Findings like the one from [1] have led to a variety of ETC schemes aiming at the reduction

☆ F. Allgöwer thanks the German Research Foundation (DFG) for support of this work within grant AL 316/13-2 – 285825138 and within the German Excellence Strategy under grant EXC-2075 – 390740016. F. Aurzada thanks the DFG for support within grant AU 370/7. For the cooperation until Jan. 2022, M.A. Lifshits thanks for the support within grant RFBR 20-51-12004.
  * Corresponding author.
    *E-mail addresses:* meister@ist.uni-stuttgart.de (D. Meister), aurzada@mathematik.tu-darmstadt.de (F. Aurzada), mikhail@lifshits.org (M.A. Lifshits), allgower@ist.uni-stuttgart.de (F. Allgöwer).

https://doi.org/10.1016/j.nahs.2024.101494
Received 19 April 2023; Received in revised form 6 March 2024; Accepted 29 March 2024
Available online 8 April 2024

of the sampling frequency while still achieving a certain control goal, such as maintaining a performance level. Note that some works examine the relationship between ETC and TTC performance under equal average triggering rates, e.g., [1,2], whereas others compare ETC and TTC average triggering rates under equal performance requirements, e.g., [3]. We would like to highlight that both perspectives are conceptually equivalent: A performance advantage of ETC over TTC under equal average triggering rates directly translates into a reduction of average triggering rates of ETC compared to TTC under equal performance requirements [1].

The reduction in "unnecessary" communication often appears as an argument for ETC also being advantageous for communication channels with limited bandwidth. The idea of using ETC to decrease shared medium utilization has been adopted from the field of networked control systems (NCS), e.g., [4,5], to the field of multi-agent systems (MAS), as seen in works such as [6,7]. To distinguish between NCS that are only coupled through their usage of a shared communication medium and MAS in which agents also cooperate to achieve a common goal, the former will be referred to as non-cooperative NCS throughout this paper.

The setup from [1] with impulsive inputs has been extended in various ways in order to find ETC schemes that are optimal with respect to a defined performance measure. Most works in this paragraph use a performance measure that is quadratic in the system state and linear in the triggering rate including a scalar trade-off factor. For first-order linear systems, [4] introduces a minimum inter-event time and aims to find the optimal triggering condition. The work [8, Paper II] establishes a closed-form solution for optimal triggering rules for the multidimensional integrator case and shows simulation-based results for the generalization to linear time-invariant systems. The authors in [9] provide a numerical design method for optimal triggering rules in an LQG setting with output feedback. Their work builds upon [10–12] which provide an $\mathcal{H}_2$-optimal controller design method for any given uniformly bounded sampling pattern in a linear system setup. Moreover, they prove that the design of optimal triggering rule and optimal controller are separable in their setting, which allows [9] to focus on the former. For discrete-time systems, [13] proposes an optimal periodic ETC design method for linear time-invariant systems.

Since finding optimal ETC schemes remains challenging, [2,14–17] have introduced and evaluated a so-called consistency property of ETC schemes in various LQ- and $\mathcal{L}_2/\ell_2$-settings. In short, an ETC scheme is considered consistent with respect to the chosen performance criterion (LQ, $\mathcal{L}_2/\ell_2$) if it guarantees the same performance level as any periodic TTC scheme while yielding a smaller (or equal) average triggering rate. An analogous definition of consistency is that the ETC scheme results in a better (or equal) performance level compared to any periodic TTC scheme while having the same average triggering rate. Thus, one can consider the work in [1] as an evaluation of LQ-consistency in a single-integrator setup with a particular choice of cost matrices. Moreover, considering the optimality perspective from the previous paragraph together with the consistency viewpoint, [3] presents an ETC design method for continuous-time LTI systems guaranteeing a certain $\mathcal{H}_\infty$-performance level. Their method co-designs the controller and the triggering rule according to the chosen $\mathcal{H}_\infty$-performance level and guarantees consistency with respect to the corresponding periodic sampled-data controller.

Following another research direction, [18] extended the results from [1] to incorporate also network effects such as packet loss in the comparison of TTC and ETC for the non-cooperative single-integrator NCS case. They point out that ETC can perform worse than TTC above a certain packet loss probability. In [19] and [20], transmission delays are included into the comparison and the packet loss probability is determined based on the medium access protocol. At last, [21] provides a performance comparison of TTC and ETC schemes for single-integrator systems considering various medium access protocols. The authors demonstrate the impact of the network load on the performance of the single-integrator NCS for various triggering schemes and medium access protocols. Thereby, they establish the importance of taking the properties of the communication network into consideration when designing triggering schemes for NCS. Another performance comparison in this realm is presented in [17] which analyzes linear discrete-time systems and contrasts purely stochastic with stochastic event-based triggering rule performance. Analyzing more general NCS and their behavior under (periodic) ETC schemes is an active field of research, e.g., [22,23].

Although some fundamental considerations have shown that TTC can sometimes outperform ETC if network effects are taken into account, ETC is still very popular for NCS. As discussed previously in this section, many settings not suffering from network losses provably yield a performance improvement under ETC when compared to TTC. This also led to various ETC approaches for MAS while there exists no work on the fundamental characteristics of TTC compared to ETC in this case. As pointed out by [7], the event-triggered consensus literature is still missing performance analyses that quantify the benefit of ETC over TTC schemes. This work aims to close this gap in order to understand whether qualitative results are the same for MAS as in the non-cooperative NCS case, or whether new effects might arise. With the previously discussed works in mind, we provide a first theoretical evaluation of ETC and TTC performance by analyzing a simple MAS problem. Specifically, we naturally extend the analysis in [1] and [8, Paper II] to a distributed consensus problem considering ETC with a level-triggering rule and assuming delay-free communication between the agents. Our main contribution is the finding that, for this setup, ETC is not always superior to TTC, even without considering packet loss or transmission delays. In particular, we show that we lose the expected performance advantage of ETC over TTC beyond a critical number of agents in the network. The existence of such a critical number of agents is of interest by itself. In addition, this article thereby demonstrates that the design of performant decentralized ETC schemes can pose additional challenges compared to the analogue non-cooperative NCS case. Furthermore, we provide the asymptotic order of the performance measure for ETC and TTC as a function of the number of agents. This gives further insights into the relationship between ETC and TTC in our particular MAS setup. Compared to the corresponding conference paper [24], we provide extensive proof details for all our results. In addition, we extend our statements to more general communication topologies than all-to-all networks under the assumption of no communication delays. Moreover, we generalize our setup to a class of provably optimal control inputs. Firstly, this allows us to show that our results hold for a broader class of problems. Secondly, further potential performance improvements via a better choice of the control input can be ruled out in our analysis. Furthermore, we improve our simulation results by increased sample numbers, simulations for more network sizes and by complementing the obtained results with a confidence interval.

Our paper is structured as follows: We start by stating some preliminaries on background knowledge, especially graph theory, and on our notation in Section 2. In Section 3, we introduce the setup and formulate the considered problem. After that, we present our theoretical results in Section 4 and demonstrate our findings in a numerical simulation in Section 5. We conclude this work in Section 6 and provide additional proofs and details in the appendix.

## 2. Preliminaries

In this section, we introduce relevant notation, especially but not exclusively regarding graph theory. A graph $\mathcal{G} = (\mathcal{V}, \mathcal{E})$ consists of a set of vertices $\mathcal{V} = \{1, \ldots, N\}$, also referred to as nodes, and a set of edges $\mathcal{E} \subset \mathcal{V} \times \mathcal{V}$. The graph $\mathcal{G}$ is called undirected if $(i, j) \in \mathcal{E} \Leftrightarrow (j, i) \in \mathcal{E}$ for all $i, j \in \mathcal{V}$. We will focus on definitions for undirected graphs throughout the rest of this section. If an edge $(i, j) \in \mathcal{E}$ exists between two nodes, they are called adjacent. All nodes that are adjacent to node $i$ are also referred to as node $i$'s neighbors $j \in \mathcal{N}_i = \{j \in \mathcal{V} \mid (i, j) \in \mathcal{E}\}$. Note that we generally exclude self-loops, i.e., $(i, i) \notin \mathcal{E}$. In the MAS context, the adjacency of nodes $i$ and $j$ indicates that agents $i$ and $j$ are able to communicate with each other. Nodes $i$ and $j$ are referred to as connected if there exists a path between those nodes, i.e., a sequence of distinct nodes, starting at $i$ and ending at $j$, such that each pair of consecutive nodes is adjacent. If all pairs of nodes in graph $\mathcal{G}$ are connected, the graph is called connected.

Furthermore, the adjacency matrix $A$ consists of elements $a_{ij} = 1$ if $i$ and $j$ are adjacent and $a_{ij} = 0$ otherwise. For an undirected graph $\mathcal{G}$, the adjacency matrix is symmetric. In addition, the degree $d_i$ of a node $i$ denotes the number of neighbors of node $i$, i.e., the cardinality of the neighbor set $|\mathcal{N}_i|$. The degree matrix $D$ of graph $\mathcal{G}$ is the diagonal matrix $D = \mathrm{diag}(d_1, \ldots, d_N)$. With the definitions of adjacency and degree matrix, the Laplacian matrix $L$ of graph $\mathcal{G}$ is defined as $L = D - A$. For an undirected graph $\mathcal{G}$, the Laplacian matrix $L$ is symmetric and positive semi-definite. Moreover, it is column and row stochastic and, thus, has the eigenvector of all ones corresponding to the eigenvalue 0. If $\mathcal{G}$ is also connected, the Laplacian matrix $L$ has exactly one zero eigenvalue. Note that we can compute the cardinality of the edge set as $|\mathcal{E}| = \mathrm{tr}(L) = \mathrm{tr}(D)$. Due to the definition via directed edges, the cardinality of $\mathcal{E}$ is twice as large as the number of undirected edges in the graph $\mathcal{G}$.

Beyond graph theory, we utilize two notation alternatives regarding the sequence of transmission events in this paper: On the one hand, we refer to the sequence of triggering time instants with the notation $(t_k^j)_{k \in \mathbb{N}}$ for agent $j \in \{1, \ldots, N\}$ where $\mathbb{N}$ represents the set of positive integers. On the other hand, we denote the event sequence of the complete MAS by $(t_k)_{k \in \mathbb{N}}$. For some formulations in this work, let us additionally define $t_0 = 0$. Naturally, ordering the event sequence $(t_k^j)_{k \in \mathbb{N}}$ for all agents $j \in \{1, \ldots, N\}$ in an increasing fashion yields the sequence $(t_k)_{k \in \mathbb{N}}$. If any elements in $(t_k^j)_{k \in \mathbb{N}}$ for all agents $j \in \{1, \ldots, N\}$ should be equal, we subsume them in a single $t_k$ in the sequence $(t_k)_{k \in \mathbb{N}}$.

Finally, let us denote the expected value and the variance of a random variable by $\mathbb{E}[\cdot]$ and $\mathbb{V}[\cdot]$, respectively. In addition, let $\delta(\cdot)$ refer to the Dirac delta impulse and $\mathbb{1}_{(\cdot)}$ denote the indicator function. Moreover, let $\mathbb{R}$ abbreviate the set of all real numbers and $\lim_{\epsilon \downarrow 0}$ indicate the right-sided limit. Lastly, $a_n \sim b_n$ means that $\lim_{n \to \infty} a_n/b_n = 1$ for arbitrary sequences $(a_n)_{n \in \mathbb{N}}$, $(b_n)_{n \in \mathbb{N}}$

## 3. Problem formulation

In this section, we introduce the considered setup and derive the optimal control input for the formulated problem.

### 3.1. Setup

We consider an MAS consisting of $N$ single-integrator agents that are perturbed by noise

$$\mathrm{d}x_i = u_i \mathrm{d}t + \mathrm{d}v_i, \tag{1}$$

starting in consensus, i.e., initial states $x_i(0) = 0$ for all $i \in \{1, \ldots, N\}$, and with $v_i(t)$ referring to a standard Brownian motion and $u_i(t)$ to the control input. Let the agents be able to communicate according to an undirected connected communication graph with $N$ nodes representing the agents and Laplacian $L$. Therefore, an agent $i$ is able to communicate with its neighbors $j \in \mathcal{N}_i$.

Furthermore, we presume that the agents can continuously monitor their own state and trigger discrete transmission events in order to share information with their neighbors. The shared information is then used to preserve consensus between the agents as well as possible. Thus, the control inputs $u_i(t)$ are required to be causal in the sense that they only depend on information transmitted up to time $t$ and local information up to the latest triggering time instant with respect to $t$. As explained in the introduction, we aim at comparing TTC and ETC schemes for triggering transmissions. For that purpose, let us consider the cost functional

$$J := \limsup_{M \to \infty} \frac{1}{M} \int_0^M \mathbb{E}\left[x(t)^\top L x(t)\right] \mathrm{d}t \tag{2}$$

as a performance measure where $x(t) = [x_1(t), \ldots, x_N(t)]^\top$. It quantifies the expected quadratic deviation from consensus and can also be written as

$$J = \limsup_{M \to \infty} \frac{1}{M} \int_0^M \mathbb{E}\left[\frac{1}{2} \sum_{(i,j) \in \mathcal{E}} (x_i(t) - x_j(t))^2\right] \mathrm{d}t.$$

**Remark 1.** The quadratic term $x^\top L x$ is a typical measure for the deviation of an MAS from consensus and, for example, also often used as a Lyapunov function, see, e.g., [25]. From an optimal control viewpoint, we consider a quadratic state cost with positive semi-definite weight matrix and no input cost.

**Remark 2.** We do not incorporate a cost term on the triggering rate in (2) since we will compare TTC and ETC under equal average triggering rates, cf. Section 4.3 and, e.g., [2]. Any cost component related to the average triggering rate can therefore be neglected for the comparison.

**Remark 3.** Note that the usual motivation behind the application of an ETC scheme compared to a TTC scheme is the capability of the former to reduce the required triggering rate while still achieving a certain control goal, such as maintaining a chosen performance level. In other words, the TTC scheme is typically expected to perform worse at the reduced triggering rate the ETC scheme has to offer. Therefore, we would like to highlight that a comparison of ETC and TTC average triggering rates under equal performance requirements is conceptually equivalent to a comparison of ETC and TTC performance under equal average triggering rates. This observation has also been utilized in [1,8] as well as [2,14–17]. It provides the motivation for the performance comparison of ETC and TTC schemes under equal average triggering rates conducted in this paper. This also follows the survey paper [26] stating that "The quantitative evaluation of all these [event- and self-triggered] strategies should reflect both control costs such as quadratic costs as in LQR control or relevant $\mathcal{L}_p$-gains, and communication costs such as average sampling rates, minimal inter-event times, or transmission power". Note that the choice of performance measure depends on the setup and should quantify control goal satisfaction in a meaningful way. In summary, we compare ETC and TTC control costs under equal communication costs in this article.

We are well aware that the considered setup is simple and does not cover the vast variety of practically relevant settings available in the literature on cooperative control. However, the simplicity of the setup allows for its detailed analysis and understanding. The motivation behind this work is to provide theoretical results for the performance comparison between TTC and ETC in such a simple cooperative setup and, thereby, uncover new phenomena and differences in the outcome when compared to existing results. For that purpose, note additionally that we study the same setup as in [1] except for the fact that we consider a cooperative control goal in a distributed setting. This will allow us to contrast the findings later on.

### 3.2. Optimal control input

Given the setup described in the previous section, we can now characterize an optimal causal control input with respect to (2). As the control input turns out to be independent of the deployed triggering scheme, we consider it to be part of the problem formulation. In particular, the following proposition establishes that an impulsive control input which resets all agents to consensus instantaneously at each triggering instant is an optimal causal controller. For computing the required control input, each agent $i$ can use local information and information received through communication up to the latest triggering instant. This renders the considered control architecture distributed according to [27,28], and we combine the respective data into locally available sets of information $\mathcal{I}_{t,i}$. Thus, at each instant in the triggering time sequence $(t_k)_{k\in\mathbb{N}}$, every agent $i$ resets to a consensus point $c_i(\mathcal{I}_{t_k,i}, t_k)$ common to all agents and, thus, satisfying $c_i(\mathcal{I}_{t_k,i}, t_k) = c_j(\mathcal{I}_{t_k,j}, t_k)$ for all $(i,j) \in \mathcal{E}$, $k \in \mathbb{N}$. Note that we will consider sequences $(t_k)_{k\in\mathbb{N}}$ consisting of symmetric stopping times in the following proposition. A stopping time is called symmetric if the sign of the governing stochastic process can be changed without influencing the stopping time itself. The triggering schemes considered in this work will yield stopping times that are indeed symmetric, as we will see in subsequent sections.

**Proposition 1.** *Let us consider triggering time sequences $(t_k)_{k\in\mathbb{N}}$ of symmetric stopping times, agent dynamics (1), and performance measure (2). Moreover, let $c_i(\mathcal{I}_{t_k,i}, t_k)$ be a common consensus point among all agents such that $c_i(\mathcal{I}_{t_k,i}, t_k) = c_j(\mathcal{I}_{t_k,j}, t_k)$ for all $(i,j) \in \mathcal{E}$, $k \in \mathbb{N}$. In addition, let $\mathcal{I}_{t,i}$ contain local and broadcast information for agent $i$ until the latest triggering instant with respect to $t$. Then, the impulsive control input $u(t) = [u_1(t), \ldots, u_N(t)]^\top$ with*

$$u_i(t) = \sum_{k \in \mathbb{N}} \delta(t - t_k)(c_i(\mathcal{I}_{t_k,i}, t_k) - x_i(t_k)),$$

*is an optimal causal control input.*

**Proof.** Can be found in Appendix A.

The control input class $u(t)$ specified in Proposition 1 resets all agents instantaneously to consensus at each triggering time instant. Between those triggering time instants, no control input is applied, and the agents behave according to standard Brownian motions. Any such resetting control input is optimal under the considered performance measure (2). The resulting consensus point $c_i(\mathcal{I}_{t_k,i}, t_k)$ is irrelevant for our analysis.

Let us give a few examples of control schemes that belong to the class in Proposition 1 to provide some intuition:

1. *All-to-all communication topology and one-to-all broadcast:* The agents are controlled by the impulsive control input

$$u_i(t) = \sum_{k \in \mathbb{N}} \sum_{j \in \mathcal{N}_i} \delta(t - t_k^j)(x_j(t_k^j) - x_i(t_k^j)), \tag{3}$$

where $\mathcal{N}_i = \{1, \ldots, N\} \setminus \{i\}$ and $t_k^j$ denotes the transmission time instant of packet $k$ from agent $j$. Thus, the system is reset to consensus by transmitting one agent's state $x_j(t_k^j)$ to all other agents. All the receiving agents then reset their state to the broadcast state $x_j(t_k^j)$. The local information set can therefore be written as $\mathcal{I}_{t,i} = \{x_i(s) \mid \exists k \in \mathbb{N} : s = t_k, s \le t\} \cup \bigcup_{j \in \mathcal{N}_i} \{x_j(s) \mid \exists k \in \mathbb{N} : s = t_k^j, s \le t\}$.

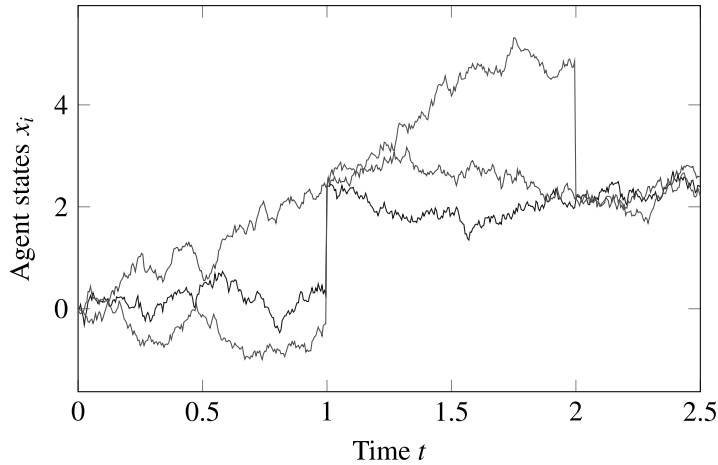

**Fig. 1.** Example for an MAS state evolution under TTC with constant inter-event time $T_{\text{TT}} = 1$ following the control scheme of Example 1.

2. *Multi-hop communication and network flooding:* The same scheme as in the previous example also works for arbitrary connected communication graphs if all agents pass on received messages to their respective neighbors. This is referred to as multi-hop communication and distributes transmitted information within the network beyond local neighbor clusters, cf. [29] for a multi-hop protocol involving continuous communication. As long as the information is spread throughout the complete network of agents, we are able to apply (3) analogously in this case: The message $x_j(t_k^j)$ is made available to all agents by means of multi-hop communication. A related method is the flooding algorithm in networks which refers to nodes passing on received information until it is known to all network participants, see, e.g., [30] and, for an advanced scheme, [31].
   Note that the proposed methods usually induce a significant communication delay. In this work, we consider an idealized setup without delays such that consensus can be achieved instantaneously. We refer the reader to [32] for a consideration of communication delays in an analysis involving complete communication topologies. Further studies on delays in this context but for other topologies are an open topic for future research.
3. *Reset to the origin:* The setup with $c_i(\mathcal{I}_{t,i}, t) = 0$ for all $i \in \{1, \ldots, N\}$ in which all agents are reset to the origin at each triggering instant also belongs to the defined control input class. In this special case, only the reset time instants need to be communicated to all agents in the network. The local information set can therefore be written as $\mathcal{I}_{t,i} = \{x_i(s) \mid \exists k \in \mathbb{N} : s = t_k, s \leq t\}$. Note that this example may seem unexciting for solving cooperative problems, but highlights the similarity to the control inputs considered in related analyses in the non-cooperative NCS literature, such as [1] or [8, Paper II].

In Fig. 1, we provide an example for a state evolution of an MAS with 3 agents under a TTC scheme. The MAS is reset to consensus at two time instants.

The specified examples are not an exhaustive list of possible schemes that belong to the described control input class. The following performance analysis holds for any scheme that fits into this class of control inputs.

**Remark 4.** Note that the performance analysis even remains the same if we have an unconnected graph since the performance measure would then also only require cluster consensus for minimal cost.

**Remark 5.** We would like to highlight that the provided control scheme examples can be realized without the requirement of time synchronization. For Example 1 under a TTC scheme, this could be realized in a leader-follower setup where it is always the same agent broadcasting its state to the others. The other agents can then react on reception of the communicated information. Similar arguments also apply to the other examples. Thus, neither TTC nor ETC require time synchronization in the provided examples which allows for a fair comparison of the schemes.

**Remark 6.** Note that control inputs from the class specified in Proposition 1 basically solve the typical consensus problem instantaneously. For that, some information typically needs to be communicated globally, as also seen in the provided examples. While this strategy is optimal as shown in Proposition 1, there might be practical constraints prohibiting the application of such a scheme, for example, input constraints or non-zero communication delays. It is therefore a meaningful research direction to explore other control strategies that can cope with such practical limitations. Moreover, analysis under such constraints may well introduce additional effects on the performance relationship of ETC and TTC, e.g., with respect to the communication topology. However, our goal in this work is to establish statements under idealized conditions such that the number of impact factors on the found results is limited. Thus, considering the class of optimal control inputs under the presented assumptions is a meaningful first step.

## 4. Main results

In this section, we introduce the two triggering schemes and derive and compare the related cost according to (2).

### 4.1. Preliminaries

Let us first establish some facts on the considered problem which we can build upon in the following analysis. Similar to [18], we find

**Lemma 1.** *Let the sequence of inter-event times be independent and identically distributed and let the expected value of the inter-event times be bounded, i.e., $\mathbb{E}[T] = \mathbb{E}[t_1] < \infty$. Then, the cost (2) is equal to*

$$J(T) = \frac{\mathbb{E}\left[\frac{1}{2} \sum_{(i,j) \in \mathcal{E}} \int_0^T \left(x_i(t) - x_j(t)\right)^2 \mathrm{d}t\right]}{\mathbb{E}[T]},$$

*where $T = t_1$ is the inter-event time determined by the respective triggering scheme.*

**Proof.** Can be found in Appendix B.

Under the assumptions of Lemma 1, we can thus compute the cost by considering only the first sampling interval. Denoting $Q(T) := \mathbb{E}\left[\frac{1}{2} \sum_{(i,j) \in \mathcal{E}} \int_0^T \left(x_i(t) - x_j(t)\right)^2 \mathrm{d}t\right]$, we can express the numerator of the cost as follows.

**Lemma 2.** *Let $T$ be a symmetric stopping time, i.e., if one replaces $v_i$ by $-v_i$ for any $i \in \{1, \ldots, N\}$ the value of $T$ does not change, as well as independent of the direction, i.e., $T$ does not change if $v_i$ is interchanged with $v_j$ for any $i, j \in \{1, \ldots, N\}$. Then, under the assumptions of Lemma 1, the following holds*

$$Q(T) = |\mathcal{E}| \cdot \mathbb{E}\left[\int_0^T v_1(t)^2 \, \mathrm{d}t\right].$$

**Proof.** We start with the expression

$$Q(T) = \mathbb{E}\left[\frac{1}{2} \sum_{(i,j) \in \mathcal{E}} \int_0^T (v_i(t) - v_j(t))^2 \, \mathrm{d}t\right] = \mathbb{E}\left[\frac{1}{2} \sum_{(i,j) \in \mathcal{E}} \int_0^T (v_i(t)^2 - 2v_i(t)v_j(t) + v_j(t)^2) \, \mathrm{d}t\right].$$

By assumption, the stopping time $T$ is symmetric. Observe that the distribution of the random variable $\int_0^T v_i(t)v_j(t) \, \mathrm{d}t$ is symmetric as well since replacing $v_i$ by $-v_i$ only changes the sign of the integrand. Therefore, the expectation of the mixed term is zero for any $i \neq j$. Thus, the following holds

$$Q(T) = \mathbb{E}\left[\int_0^T \sum_{\substack{(i,j) \in \mathcal{E}: \\ i < j}} (v_i(t)^2 + v_j(t)^2) \, \mathrm{d}t\right] = \mathbb{E}\left[\int_0^T \sum_{i=1}^N d_i v_i(t)^2 \, \mathrm{d}t\right] = |\mathcal{E}| \cdot \mathbb{E}\left[\int_0^T v_1(t)^2 \, \mathrm{d}t\right],$$

using that $T$ is independent of the direction. □

### 4.2. Time-triggered control

As a comparison benchmark for ETC, we choose a TTC scheme in which the transmission events are scheduled periodically with a constant inter-event time $T_{\mathrm{TT}} = t_{k+1} - t_k = $ const. for all $k \in \mathbb{N}_0$. This is in line with the consistency definition for ETC introduced and considered in [2,14–17]. Given the first setup example from Section 3.2, an all-to-all communication topology and one-to-all broadcast, this implies that the transmission of one agent's state to all the others takes place with a fixed frequency. This state information is then used by all other agents to reset their states to consensus. How the transmitting agent is chosen in this particular setting plays no role for the performance analysis to come. An example for an MAS state evolution under TTC is shown in Fig. 1. Deploying this triggering scheme in the considered setup leads to the following theorem.

**Theorem 1.** *Suppose agents with dynamics (1) are controlled by an impulsive input from the class in Proposition 1 with constant inter-event times $T_{\mathrm{TT}}$. Then, the cost (2) is given by*

$$J_{\mathrm{TT}}(T_{\mathrm{TT}}) = |\mathcal{E}| \cdot \frac{T_{\mathrm{TT}}}{2}.$$

**Proof.** Since the inter-event times $T_{\mathrm{TT}}$ are identical and constant, it suffices to analyze the interval between two transmissions and Lemmas 1 and 2 apply. Thus, we can write (2) as $J_{\mathrm{TT}}(T_{\mathrm{TT}}) = Q(T_{\mathrm{TT}})/T_{\mathrm{TT}}$ with

$$Q(T_{\mathrm{TT}}) = |\mathcal{E}| \cdot \int_0^{T_{\mathrm{TT}}} \mathbb{E}[v_1(t)^2] \, \mathrm{d}t = |\mathcal{E}| \cdot \int_0^{T_{\mathrm{TT}}} t \, \mathrm{d}t = |\mathcal{E}| \cdot \frac{T_{\mathrm{TT}}^2}{2},$$

as required. □

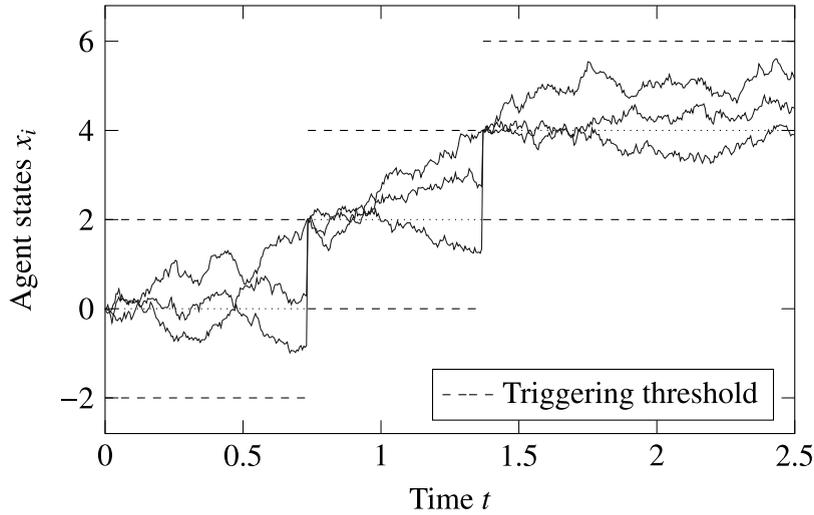

**Fig. 2.** Example for an MAS state evolution under ETC with $\Delta = 2$.

**Remark 7.** The result for the cost in the TTC case is the same as in [1] but scaled by twice the number of connected agent pairs $|\mathcal{E}|$. This is also related to the results for non-cooperative NCS in [19] and related papers where the cost scales with the number of network participants $N$.

*4.3. Event-triggered control*

In ETC, the necessity to communicate is captured by a continuously evaluated triggering condition. Once the condition is fulfilled, a transmission event is initiated by the respective agent. Since we are operating in a distributed setting, each agent evaluates its triggering condition locally. Consequently, only local information is to be used in the respective triggering rule. This is in line with the definition of decentralized event-triggered consensus in [7, Problem 2.5]. We are thus considering a distributed control architecture according to [27,28], as explained in Section 3.2, and deploy decentralized triggering schemes in line with the definition [7, Problem 2.5]. This combination is what we refer to when utilizing the terminology "distributed" for our setting in this article.

Since the agents are for example incorporating local state information in the triggering decision, ETC is often argued to lower the communication rate while maintaining the same performance level as TTC, see, e.g., [6]. For this work, we use

$$|x_i(t) - x_i(t_{\hat{k}(t)})| \geq \Delta \tag{4}$$

as the triggering condition where $\hat{k}(t) = \max \{k \in \mathbb{N}_0 \mid t_k \leq t\}$ and $\Delta > 0$. It compares the local state deviation from the state at the latest event $x_i(t_{\hat{k}(t)})$ to a threshold $\Delta$. This form of triggering rule is quite common in distributed setups, see, for example, [25], and also referred to as level-triggering rule.

**Remark 8.** Note that we use a triggering condition that is analogous to the one in [1,18,21]. The analogy between the problem setups here and in the respective works motivates this choice of triggering rule since we will contrast the findings in this article. For transient problems, this rule is known to yield only practical stability of the closed loop. However, we study a problem in which we aim for maintaining the consensus state while the agent are perturbed by noise. In analogy to [1,8], the level-triggering rule is therefore a meaningful choice for the considered setting.

**Remark 9.** We choose the same threshold $\Delta$ for all agents. Note that strictly speaking this is only the best choice if the contribution of each agent's state to the cost is equal, namely if all agents $i$ have the same degree $d_i$. For heterogeneous degrees $d_i$, a heterogeneous choice of $\Delta_i$ might be advantageous. Deriving the optimal choice for $\Delta_i$ in this case is beyond the scope of this paper and, thus, the analysis for heterogeneous $\Delta_i$ is not considered in the remainder of this paper.

Considering the first setup example from Section 3.2 with an all-to-all communication topology and one-to-all broadcast, the described ETC scheme leads to one agent broadcasting its state to the others once the respective local triggering condition is fulfilled. As in the time-triggered case, this state information is then used by all other agents to reset their states to consensus. The triggering agent is thus chosen as the transmitting agent resulting in a decentralized scheme. An example for an MAS state evolution under the analyzed ETC scheme is shown in Fig. 2.

Utilizing Lemmas 1 and 2 again allows us to analyze the cost on the first sampling interval, also in the ETC case. In contrast to the TTC analysis, the length of this time interval is described by a probabilistic stopping time $T_{\text{ET}}(\Delta) = \inf\{t > 0 \mid \exists i \in \{1, \ldots, N\} : |x_i(t)| = \Delta\}$. Although we are not able to derive an explicit expression for the cost $J_{\text{ET}}(\Delta) := J(T_{\text{ET}}(\Delta))$ for the ETC case, we can still arrive at results on its relationship to the TTC cost $J_{\mathrm{TT}}(T_{\mathrm{TT}})$ derived in Section 4.2. Note that the latter relationship is also what we are primarily interested in for this work.

In order to facilitate a fair comparison between $J_{\mathrm{ET}}(\Delta)$ and $J_{\mathrm{TT}}(T_{\mathrm{TT}})$, we require $T_{\mathrm{TT}} = \mathbb{E}[T_{\mathrm{ET}}(\Delta)]$ which results in the same average triggering rate for both schemes. This is again inspired by the line of thought for the consistency property of ETC schemes considered in [2,14–17]. Note that this constraint embodies the bridge between the triggering threshold $\Delta$ determining $\mathbb{E}[T_{\mathrm{ET}}(\Delta)]$ for the ETC scheme and the constant inter-event time $T_{\mathrm{TT}}$ in the TTC case.

Let us first establish the following lemmas, which will enable us to focus on the case $\Delta = 1$ for the derivations to come.

**Lemma 3.** *Under triggering condition* (4), *the following scaling relationships hold*

$$Q_{\mathrm{ET}}(\Delta) = \Delta^4 Q_{\mathrm{ET}}(1), \qquad \mathbb{E}[T_{\mathrm{ET}}(\Delta)] = \Delta^2 \mathbb{E}[T_{\mathrm{ET}}(1)], \qquad \mathbb{V}[T_{\mathrm{ET}}(\Delta)] = \Delta^4 \mathbb{V}[T_{\mathrm{ET}}(1)],$$

*where* $Q_{\mathrm{ET}}(\Delta) := Q(T_{\mathrm{ET}}(\Delta))$.

**Proof.** Let us show the first equality. Note that the assumptions of Lemmas 1 and 2 are fulfilled under triggering condition (4) with $T_{\mathrm{ET}}(\Delta) = \inf\{t > 0 \mid \exists i \in \{1, \ldots, N\} : |x_i(t)| = \Delta\}$. Therefore, we have

$$\begin{aligned}
Q_{\mathrm{ET}}(\Delta) &= |\mathcal{E}| \cdot \mathbb{E}\left[\int_0^{T_{\mathrm{ET}}(\Delta)} v_1(s)^2 \, \mathrm{d}s\right] \\
&= |\mathcal{E}| \cdot \mathbb{E}\left[\int_0^{\inf\{t>0 \mid \exists k : |v_k(t)| = \Delta\}} v_1(s)^2 \, \mathrm{d}s\right] \\
&= |\mathcal{E}| \cdot \mathbb{E}\left[\int_0^{\inf\{t>0 \mid \exists k : |v_k(\Delta^2 t/\Delta^2)| = \Delta\}} v_1(\Delta^2 s/\Delta^2)^2 \, \mathrm{d}s\right] \\
&= |\mathcal{E}| \cdot \mathbb{E}\left[\int_0^{\inf\{t>0 \mid \exists k : \Delta |v_k(t/\Delta^2)| = \Delta\}} \Delta^2 v_1(s/\Delta^2)^2 \, \mathrm{d}s\right] \\
&= \Delta^2 |\mathcal{E}| \cdot \mathbb{E}\left[\int_0^{\Delta^{-2} \inf\{\Delta^2 t' > 0 \mid \exists k : |v_k(t')| = 1\}} v_1(s')^2 \Delta^2 \, \mathrm{d}s'\right] \\
&= \Delta^4 |\mathcal{E}| \cdot \mathbb{E}\left[\int_0^{\inf\{t' > 0 \mid \exists k : |v_k(t')| = 1\}} v_1(s')^2 \, \mathrm{d}s'\right] \\
&= \Delta^4 Q_{\mathrm{ET}}(1).
\end{aligned}$$

In the fourth step, we used the scaling property of Brownian motions and, in the fifth step, we applied linear integral substitution. All other formulas are proved similarly. □

Thus, we can derive relevant quantities for the considered setup for $\Delta = 1$ and use Lemma 3 to generalize the found expressions to arbitrary choices of $\Delta$. Moreover, we obtain the following lemma as a direct consequence of Lemma 3.

**Lemma 4.** *Let* $J_{\mathrm{TT}}(\mathbb{E}[T_{\mathrm{ET}}(\Delta)])$ *denote the cost under constant inter-event times* $T_{\mathrm{TT}} = \mathbb{E}[T_{\mathrm{ET}}(\Delta)]$. *Then, the following holds*

$$\frac{J_{\mathrm{ET}}(\Delta)}{J_{\mathrm{TT}}(\mathbb{E}[T_{\mathrm{ET}}(\Delta)])} = \frac{J_{\mathrm{ET}}(1)}{J_{\mathrm{TT}}(\mathbb{E}[T_{\mathrm{ET}}(1)])}.$$

**Proof.** Due to Lemma 3 together with Lemma 1 and Theorem 1, we have

$$J_{\mathrm{ET}}(\Delta) = \Delta^2 J_{\mathrm{ET}}(1),$$
$$J_{\mathrm{TT}}(\mathbb{E}[T_{\mathrm{ET}}(\Delta)]) = \Delta^2 J_{\mathrm{TT}}(\mathbb{E}[T_{\mathrm{ET}}(1)]).$$

Computing the ratio $J_{\mathrm{ET}}(\Delta)/J_{\mathrm{TT}}(\mathbb{E}[T_{\mathrm{ET}}(\Delta)])$ with these equalities yields the desired result. □

**Remark 10.** An interpretation of Lemma 4 is that the cost comparison between $J_{\mathrm{ET}}(\Delta)$ and $J_{\mathrm{TT}}(\mathbb{E}[T_{\mathrm{ET}}(\Delta)])$ is not influenced by the choice of $\Delta$. In particular, this implies that having $\Delta$ depend on the number of agents $N$ or any tuning of $\Delta$ does not change the cost comparison results.

In summary, Lemmas 3 and 4 allow us to concentrate on the case $\Delta = 1$ for the remainder of this section. Moreover, they enable us to focus on $\Delta = 1$ in the simulation in Section 5. Before arriving at the main result of this section, we need to characterize the asymptotic order of the moments of $T_{\mathrm{ET}}(1)$.

**Lemma 5.** *Under triggering condition* (4) *with* $T_{\mathrm{ET}}(\Delta) = \inf\{t > 0 \mid \exists i \in \{1, \ldots, N\} : |x_i(t)| = \Delta\}$, *we have*

$$\mathbb{E}[T_{\mathrm{ET}}(1)] \sim \frac{1}{2 \ln N}, \tag{5}$$

$$\mathbb{E}\left[T_{\mathrm{ET}}(1)^2\right] \sim \frac{1}{(2\ln N)^2}, \tag{6}$$

$$\mathbb{V}\left[T_{\mathrm{ET}}(1)\right] \sim \frac{\pi^2/24}{(\ln N)^4}, \tag{7}$$

*where the left-hand sides implicitly depend on $N$.*

**Proof.** Throughout the proof, we drop the arguments indicating $\Delta = 1$ to simplify notation. Let $T_j := \inf\{t > 0 : |x_j(t)| = 1\}$ for all $j \in \{1, \ldots, N\}$ and, thus, $T_{\mathrm{ET}} = \inf_{1 \leq j \leq N} T_j$. Using the tail behavior derived from [33], Theorem 7.45,

$$\mathbb{P}(T_j \leq w) = \mathbb{P}(\sup_{0 \leq t \leq w} |v_j(t)| \geq 1) = \mathbb{P}(\sup_{0 \leq t \leq 1} |v_j(t)| \geq w^{-1/2}) \overset{w \to 0}{\sim} \frac{\kappa}{w^{-1/2}} \exp(-w^{-1}/2),$$

for $\kappa = \sqrt{2/\pi}$, and the independence of the exit times $T_j$, one can derive the limit theorem

$$2(\ln N)^2 \left(T_{\mathrm{ET}} - a_N\right) \Rightarrow G, \quad \text{as } N \to \infty, \tag{8}$$

with

$$a_N := \frac{1}{2\ln N} - \frac{\ln \frac{\kappa}{(2\ln N)^{1/2}}}{2(\ln N)^2},$$

and where $\Rightarrow$ stands for convergence in distribution. Moreover, $G$ is a Gumbel-distributed random variable,

$$\mathbb{P}(G \geq r) = \exp(-\exp(r)).$$

Eq. (8) can be derived from [34], Theorem 2.1.6. A direct proof is given here: Indeed, for any $r \in \mathbb{R}$, we have

$$\mathbb{P}(2(\ln N)^2\left(T_{\mathrm{ET}} - a_N\right) \geq r) = \mathbb{P}(T_{\mathrm{ET}} \geq \frac{r}{2(\ln N)^2} + a_N) = \mathbb{P}(\forall j = 1, \ldots, N : T_j \geq \frac{r}{2(\ln N)^2} + a_N)$$

$$= \mathbb{P}(T_1 \geq \frac{r}{2(\ln N)^2} + a_N)^N = \left(1 - \mathbb{P}(T_1 < \frac{r}{2(\ln N)^2} + a_N)\right)^N$$

$$\sim \left(1 - \frac{\kappa}{c_N} \exp\left(-\frac{1}{2}\left(\frac{r - \ln \frac{\kappa}{c_N}}{2(\ln N)^2} + \frac{1}{2\ln N}\right)^{-1}\right)\right)^N$$

$$= \left(1 - \frac{\kappa}{c_N} \exp\left(-\ln N \left(\frac{r - \ln \frac{\kappa}{(2\ln N)^{1/2}}}{\ln N} + 1\right)^{-1}\right)\right)^N$$

$$\sim \left(1 - \frac{\kappa}{c_N} \exp\left(-\ln N \left(1 - \frac{r - \ln \frac{\kappa}{(2\ln N)^{1/2}}}{\ln N}\right)\right)\right)^N$$

$$= \left(1 - \frac{\kappa}{c_N} \frac{1}{N} \exp\left(r - \ln \frac{\kappa}{c_N}\right)\right)^N = \left(1 - \frac{1}{N} \exp(r)\right)^N \sim e^{-e^r},$$

as required and with $c_N = (2\ln N)^{1/2}$.

The limit theorem (8) is accompanied by the convergence of the first and second moment. The proof for this is provided in Appendix C to allow for a more concise presentation of the results. It builds upon Lebesgue's dominated convergence theorem where we need to show that $\mathbb{P}(2(\ln N)^2\left(T_{\mathrm{ET}} - a_N\right) \geq r)$ and $2r\mathbb{P}(2(\ln N)^2\left(T_{\mathrm{ET}} - a_N\right) \geq r)$ are upper bounded by integrable functions.

Taking expectations in (8) gives

$$2(\ln N)^2(\mathbb{E}\left[T_{\mathrm{ET}}\right] - a_N) \to \mathbb{E}[G].$$

Thus, the following holds

$$\mathbb{E}\left[T_{\mathrm{ET}}\right] = a_N + \frac{\mathbb{E}[G]}{2(\ln N)^2}(1 + o(1)) \tag{9}$$

$$= \frac{1}{2\ln N} + \mathcal{O}\left(\frac{\ln \ln N}{(\ln N)^2}\right).$$

Similarly, taking second moments in (8) gives

$$4(\ln N)^4 \mathbb{E}\left[(T_{\mathrm{ET}} - a_N)^2\right] \to \mathbb{E}\left[G^2\right],$$

and, hence,

$$\mathbb{E}\left[T_{\mathrm{ET}}^2\right] - 2a_N \mathbb{E}\left[T_{\mathrm{ET}}\right] + a_N^2 = \frac{\mathbb{E}\left[G^2\right]}{4(\ln N)^4}(1 + o(1)),$$

which, together with (9), yields

$$\mathbb{E}\left[T_{\mathrm{ET}}^2\right] = a_N^2 + 2a_N \frac{\mathbb{E}[G]}{2(\ln N)^2}(1 + o(1)) + \frac{\mathbb{E}\left[G^2\right]}{4(\ln N)^4}(1 + o(1)) = \frac{1}{4(\ln N)^2} + \mathcal{O}\left(\frac{1}{(\ln N)^3}\right).$$

Finally, the limit theorem can be re-written as

$$2(\ln N)^2(T_{\mathrm{ET}} - \mathbb{E}[T_{\mathrm{ET}}]) + 2(\ln N)^2(\mathbb{E}[T_{\mathrm{ET}}] - a_N) \Rightarrow G.$$

Squaring, taking expectations, and dividing by $4(\ln N)^4$ gives

$$\mathbb{E}\big[(T_{\mathrm{ET}} - \mathbb{E}[T_{\mathrm{ET}}])^2\big] + (\mathbb{E}[T_{\mathrm{ET}}] - a_N)^2 = \frac{\mathbb{E}[G^2]}{4(\ln N)^4}(1 + o(1)).$$

This implies

$$\mathbb{V}[T_{\mathrm{ET}}] = \frac{\mathbb{E}[G^2]}{4(\ln N)^4}(1 + o(1)) - (\mathbb{E}[T_{\mathrm{ET}}] - a_N)^2 = \frac{\mathbb{E}[G^2]}{4(\ln N)^4}(1 + o(1)) - \frac{\mathbb{E}[G]^2}{4(\ln N)^4}(1 + o(1)) = \frac{\mathbb{V}[G]}{4(\ln N)^4}(1 + o(1)),$$

which proves (7) because $\mathbb{V}[G] = \pi^2/6$. □

With this result, we have also shown a logarithmic dependence of the moments of $T_{\mathrm{ET}}(\Delta)$ on the number of agents. This is a crucial difference to the non-cooperative NCS case, e.g., studied in [1,18,21], and caused by the distributed nature of the considered problem. Leveraging the previous lemma, we arrive at the following main theorem.

**Theorem 2.** *Suppose agents with dynamics (1) are controlled by an impulsive input from the class in Proposition 1 with inter-event times $T_{\mathrm{ET}}(\Delta) = \inf\{t > 0 \mid \exists i \in \{1, \ldots, N\} : |x_i(t)| = \Delta\}$. Then, there exists an $N_0$ such that for all $N \geq N_0$, we have*

$$J_{\mathrm{ET}}(\Delta) > J_{\mathrm{TT}}(\mathbb{E}[T_{\mathrm{ET}}(\Delta)]),$$

*i.e., TTC outperforms ETC for all $N \geq N_0$ under equal average triggering rates.*

**Proof.** Due to Lemma 4, we can concentrate on the case $\Delta = 1$. Thus, we again use simplified notation and do not state $\Delta = 1$ explicitly in our formulas. As before, let $T_j := \inf\{t > 0 : |x_j(t)| = 1\}$ for all $j \in \{1, \ldots, N\}$ and, thus, $T_{\mathrm{ET}} = \inf_{1 \leq j \leq N} T_j$. Moreover, let $\tau := \inf_{2 \leq j \leq N} T_j \geq T_{\mathrm{ET}}$.

The key estimate is

$$\int_0^{T_{\mathrm{ET}}} v_1(t)^2 \, dt \geq \int_0^{\tau} v_1(t)^2 \, dt \, (1 - \mathbb{1}_{\tau \neq T_{\mathrm{ET}}}). \tag{10}$$

Let us evaluate the expectations. By independence of $\tau$ and $v_1$, we have

$$\mathbb{E}\left[\int_0^{\tau} v_1(t)^2 \, dt\right] = \int_0^{\infty} \mathbb{E}\big[\mathbb{1}_{t \leq \tau} v_1(t)^2\big] \, dt = \int_0^{\infty} \mathbb{E}[\mathbb{1}_{t \leq \tau}]\, \mathbb{E}[v_1(t)^2] \, dt = \mathbb{E}\left[\int_0^{\tau} t \, dt\right] = \frac{\mathbb{E}[\tau^2]}{2} > \frac{\mathbb{E}[T_{\mathrm{ET}}^2]}{2}. \tag{11}$$

Note that $v_1$ is a free Brownian motion. In particular, the dependence of the overall process on the conditioning on $|x_i(t)| \leq \Delta$ for all or some $i$ is only introduced via the quantities $T_{\mathrm{ET}}$ and $\tau$. Moreover, we do not consider such conditioning in $\mathbb{E}[v_1(t)^2]$ in (11). Thus, we can use $\mathbb{E}[v_1(t)^2] = t$.

In addition, since $\tau \leq T_2$ and using the Cauchy–Schwarz inequality, we have

$$\mathbb{E}\left[\int_0^{\tau} v_1(t)^2 \, dt \cdot \mathbb{1}_{\tau \neq T_{\mathrm{ET}}}\right] \leq \mathbb{E}\left[\int_0^{T_2} v_1(t)^2 \, dt \cdot \mathbb{1}_{\tau \neq T_{\mathrm{ET}}}\right] \leq \mathbb{E}\left[\left(\int_0^{T_2} v_1(t)^2 \, dt\right)^2\right]^{1/2} \cdot \mathbb{E}\big[\mathbb{1}_{\tau \neq T_{\mathrm{ET}}}^2\big]^{1/2}$$
$$= C \cdot \mathbb{P}(\tau \neq T_{\mathrm{ET}})^{1/2} = C \, N^{-1/2}, \tag{12}$$

where $C = \mathbb{E}[(\int_0^{T_2} v_1(t)^2 \, dt)^2]^{1/2}$ does not depend on the number of agents $N$. The last step holds because $\tau \neq T_{\mathrm{ET}}$ if and only if the process $v_1$ is the first to exit $[-1, 1]$. By symmetry, this has probability equal to $1/N$. Putting (10), (11), and (12) together, we obtain

$$\mathbb{E}\left[\int_0^{T_{\mathrm{ET}}} v_1(t)^2 \, dt\right] > \frac{\mathbb{E}[T_{\mathrm{ET}}^2]}{2} - C \, N^{-1/2} = \frac{\mathbb{E}[T_{\mathrm{ET}}]^2}{2} + \frac{\mathbb{V}[T_{\mathrm{ET}}]}{2} - C \, N^{-1/2}. \tag{13}$$

Next, the definition of the limit shows that (7) implies

$$\frac{\mathbb{V}[T_{\mathrm{ET}}]}{(\pi^2/24)/(\ln N)^4} > \frac{1}{2}$$

for all $N \geq N_1$. Furthermore, let $N_2$ be such that $\frac{1}{4} \cdot \frac{\pi^2/24}{(\ln N)^4} - C N^{-1/2} > 0$ for all $N \geq N_2$ and set $N_0 := \max(N_1, N_2)$. Note that the existence of $N_2$ can, for example, be shown by establishing divergence of the sequence $z_n = \sqrt{n}/\ln(n)^4$ and utilizing that $C$ is bounded.

Plugging the inequalities into (13), we see that for $N \geq N_0$

$$\frac{1}{|\mathcal{E}|} Q_{\mathrm{ET}} = \mathbb{E}\left[\int_0^{T_{\mathrm{ET}}} v_1(t)^2 \, dt\right] > \frac{\mathbb{E}[T_{\mathrm{ET}}]^2}{2} + \frac{1}{4} \cdot \frac{\pi^2/24}{(\ln N)^4} - C N^{-1/2} > \frac{\mathbb{E}[T_{\mathrm{ET}}]^2}{2} = \frac{1}{|\mathcal{E}|} Q(T_{\mathrm{TT}} = \mathbb{E}[T_{\mathrm{ET}}]),$$

where we also used Lemma 2 in the first step and Theorem 1 in the last step. Multiplying both sides with $|\mathcal{E}|/\mathbb{E}[T_{\mathrm{ET}}]$ gives the desired inequality. □

Thus, we have proved that ETC is not necessarily outperforming TTC if we consider a distributed setup with cooperative agents. The common result supported by works like [1], that ETC schemes outperform TTC, can therefore not be simply transferred to this distributed setting. In particular, transferring an ETC scheme that provably outperforms TTC in the non-cooperative NCS case [1,8] to an analogous MAS problem can lead to a loss of that performance advantage.

We have also shown that this finding holds for any undirected connected communication topology in the considered setting. Note that the independence from the topology is rooted in the infinitely fast control capabilities by impulsive control inputs and instantaneous communication. The communication topology might indeed play a role under more realistic assumptions like constrained inputs or non-zero communication delays.

**Remark 11.** Due to Lemma 4 and similar to its interpretation in Remark 10, it is noteworthy that the critical number of agents $N_0$ in Theorem 2 is the same for all $\Delta > 0$. It is therefore not possible to influence the performance relationship of TTC and ETC by parameter tuning in the triggering rule (4) in this particular setting.

Building upon the results so far, we are also able to characterize the asymptotic order of the performance measure in the following corollary.

**Corollary 1.** *The asymptotic order of the cost* (2) *as a function of the number of agents under both triggering schemes can be expressed as*

$$J_{\text{ET}}(1) \sim J_{\text{TT}}(\mathbb{E}[T_{\text{ET}}(1)]) \sim \frac{|\mathcal{E}|}{4 \ln N}.$$

**Proof.** Let us again omit the arguments indicating $\Delta = 1$. Utilizing Theorem 1 and plugging in (5) shows the relationship for $J_{\text{TT}}(\mathbb{E}[T_{\text{ET}}])$.

The lower bound for $J_{\text{ET}}$ follows from Theorem 2. For the upper bound, observe that

$$\mathbb{E}\left[\int_0^{T_{\text{ET}}} v_1(t)^2 \, dt\right] \leq \mathbb{E}\left[\int_0^{\tau} v_1(t)^2 \, dt\right] = \frac{\mathbb{E}[\tau^2]}{2} \sim \frac{1}{2(2\ln(N-1))^2} \sim \frac{1}{2(2\ln N)^2},$$

where we used the notation from the last proof and the fact that $\tau$ has the same distribution as $T_{\text{ET}}$ for the dimension $N-1$. Utilizing $J_{\text{ET}} = |\mathcal{E}| \cdot \mathbb{E}[\int_0^{T_{\text{ET}}} v_1(t)^2 \, dt] / \mathbb{E}[T_{\text{ET}}]$ yields the desired result. □

Thus, on the one hand, the cost for ETC and TTC grows with the same order for large numbers of agents. On the other hand, this does not imply that the difference between $J_{\text{ET}}(\Delta)$ and $J_{\text{TT}}(\mathbb{E}[T_{\text{ET}}(\Delta)])$ vanishes for large $N$.

Given the results in this section, especially Theorem 2, we can thus conclude that, in the considered distributed setting, the ETC scheme is outperformed by the TTC scheme for large numbers of agents and equal average triggering rates. This result stands in contrast to the findings of similar analyses for non-cooperative setups under the assumption of delay- and loss-free communication such as [1]. Moreover, the relationship between TTC and ETC performance might well depend on the number of agents or network participants $N$ when considering cooperative settings. At this point, we would like to highlight once more that the considered system setup is inspired by [1] and [8, Paper II]: single-integrator systems, Brownian motion noise, a quadratic average infinite horizon cost, delay-free communication, and a level-triggering rule for the ETC scheme, to name some common aspects. Thus, only the setup adaptions necessary for moving to a distributed consensus problem induce the change in the ETC and TTC performance relationship that was revealed in this section. The precise root of the phenomenon found in our work is a topic of ongoing research. An analysis on a related non-cooperative problem studying the performance implications of different $p$-norms in the level-triggering rule indicates that the root can be found in the decentralized nature of the triggering rule and not so much in the cooperative control goal itself [35]. This points into the direction that the lack of global information for the triggering decision might well be the cause of the found performance degradation of ETC compared to TTC. Further examination of the phenomenon's root is a very relevant direction for future research.

While it remains unclear how the established results generalize to other distributed problems, they still point out that performance advantages of ETC can behave differently in some distributed settings when compared to their non-cooperative counterpart. This work can therefore serve as a starting point for a careful evaluation of ETC performance for a wider range of distributed settings including more general system dynamics, other noise classes, other event-triggering conditions, etc. In this context, we refer the reader to [32] for a similar analysis restricted to complete communication topologies, but incorporating network effects like packet loss and transmission delays as well as an explicit consideration of the medium access control layer.

Note also that ETC schemes in general offer advantages beyond lower average triggering rates, e.g., their natural formulation as asynchronous control schemes or their reactive behavior when unexpected disturbances occur during run time. Our main contribution is to evaluate one important aspect among these reasons for applying ETC schemes and to reveal an unexpected outcome with respect to existing literature. We would like to highlight that ETC might very well offer performance advantages over TTC in other distributed setups. The mere existence of a disadvantageous setting with respect to ETC versus TTC can serve as a motivation to search for circumstances under which ETC provably provides performance advantages in distributed setups. Moreover, we have demonstrated that deploying a performant triggering rule from the non-cooperative NCS case [1] in a distributed problem can lead to a loss of the performance advantage over TTC. Thereby, this work points out that designing performant ETC schemes can require efforts beyond transferring schemes from the non-cooperative NCS literature. Consequently, our intention is to stimulate further research in the field of ETC for distributed systems by providing new phenomenological insights within this domain.

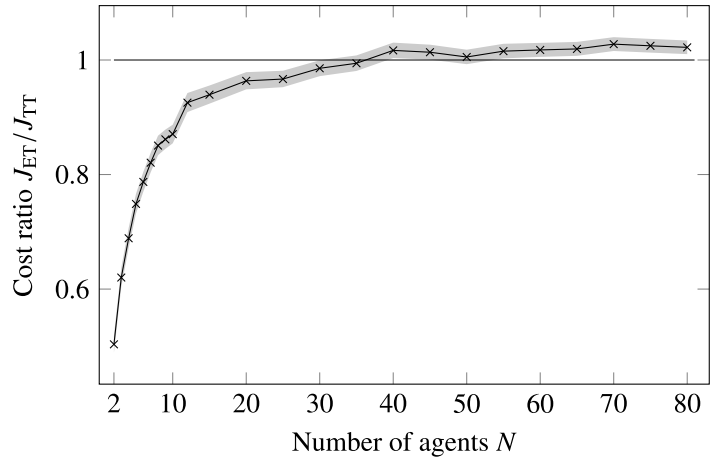

**Fig. 3.** Cost ratio of ETC over TTC based on the performed simulations.

## 5. Simulation

In this section, we perform simulations to support our theoretical findings. A simulative performance comparison involving ETC schemes is in general challenging since closed-form expressions for the expected inter-event time are rarely obtainable. Therefore, the constraint of equal expected inter-event times for different triggering schemes is hard to enforce exactly in simulation. In our case though, we can resolve this problem by estimating the expected inter-event time of the ETC scheme from simulation results and then using this estimate as inter-event time in the closed-form expression for the cost of the TTC scheme. Consequently, we are able to enforce the constraint $T_{TT} = \mathbb{E}\left[T_{ET}(\Delta)\right]$ in our setup when comparing costs of the triggering schemes.

Thus, we simulate the described MAS including the impulsive control law under the ETC scheme. With the aforementioned strategy, this allows us to estimate the cost ratio $J_{ET}(\Delta)/J_{TT}(\mathbb{E}\left[T_{ET}(\Delta)\right])$ for a varying number of agents $N$ and thereby relate the two schemes in terms of performance. To be precise, we can use the simulation estimate of $\mathbb{E}\left[T_{ET}(\Delta)\right]$ for a given $N$ with the result from Theorem 1 to compute $J_{TT}(\mathbb{E}\left[T_{ET}(\Delta)\right])$ exactly. Therefore, only the cost $J_{ET}(\Delta)$ needs to be estimated based on simulation results. We refer the reader to Appendix D for additional information on how to achieve this.

For the simulation, we set $\Delta = 1$. We can do so without loss of generality due to Lemma 4. Moreover, we simulate the MAS with the Euler–Maruyama method for $N \in \{2, 3, \ldots, 9, 10, 12, 15, 20, 25, \ldots, 80\}$ with a step size of $10^{-4}$ s. For each $N$, we perform 10 000 Monte Carlo runs for estimating $\mathbb{E}\left[T_{ET}(1)\right]$ and 250 000 Monte Carlo runs for estimating $Q_{ET}(1)$. As an analysis of the first sampling interval suffices for the cost, we can terminate a Monte Carlo run when the first event occurs. The resulting cost ratios with a minimum 95%-confidence interval are shown in Fig. 3. The derivations regarding the confidence interval can also be found in Appendix D.

As predicted by our theoretical results, we find that the ETC scheme is outperformed by TTC for larger numbers of agents. For low numbers of agents $N$, we observe a clear performance advantage of the ETC scheme in simulation. Note that this advantage is not guaranteed by the theoretical findings in this paper. In addition, the simulation results indicate that the critical number of agents $N_0$ from Theorem 2 is likely between 30 and 55 agents, also supported by the depicted confidence intervals. Thus, for this particular setting, the critical number of agents $N_0$ beyond which the TTC scheme outperforms the ETC scheme might well lie in a practically relevant range.

## 6. Conclusion

In this work, we examined TTC and ETC performance in an MAS consensus setup with single-integrator agents, optimal control inputs, and arbitrary undirected connected communication topologies. We consider an ETC scheme with a level-triggering rule, inspired by [1] and [8, Paper II]. For this particular setting, we provided a complete proof that TTC outperforms ETC beyond a certain number of agents given any communication topology in the considered class. This is in striking contrast to the outcome of similar analyses for non-cooperative NCS, as for example the ones mentioned above. In addition, we characterized the asymptotic order of the performance measure in the number of agents and evaluated our results in a numerical simulation.

This work points out that consistency considerations for ETC of MAS can be influenced by additional factors when compared to the non-cooperative NCS case, especially [1] and [8, Paper II]. In particular, performance advantages of ETC over TTC might provably vanish in a distributed setting if the number of agents is large enough. While we have shown that for a particular setup which has been chosen analogously to existing analyses in the non-cooperative NCS case, it remains to be explored how these findings generalize to other settings including different agent dynamics, noise types, event-triggering conditions, and so forth. We suggest that the transfer of experience from the non-cooperative NCS to the MAS field and the creation of new event-triggering schemes should therefore be accompanied by a careful consideration of the impact of the number of agents on the performance relationship between TTC and ETC. Moreover, this article demonstrates that the design of performant ETC schemes can pose additional challenges and is therefore a promising research direction complementing the existing non-cooperative NCS literature.

In future work, we plan to investigate the root of the discovered performance disadvantage of ETC compared to TTC for sufficiently large agent numbers more closely. Only this will allow examining and arguing about options to overcome or alleviate the found phenomenon. This can also be linked to finding a consistent ETC scheme in the sense of [15] for the considered setting. Moreover, the analysis of other setups and ETC strategies is a promising future research direction. This may also include the examination of settings with state constraints. Lastly, studying the performance implications of different communication topologies in a setting with non-zero communication delays is an interesting open research direction.

**CRediT authorship contribution statement**

**David Meister:** Writing – review & editing, Writing – original draft, Validation, Software, Investigation, Formal analysis, Conceptualization. **Frank Aurzada:** Writing – review & editing, Validation, Investigation, Formal analysis, Conceptualization. **Mikhail A. Lifshits:** Writing – review & editing, Validation, Investigation, Formal analysis. **Frank Allgöwer:** Writing – review & editing, Validation, Supervision, Conceptualization.

**Declaration of competing interest**

The authors declare that they have no known competing financial interests or personal relationships that could have appeared to influence the work reported in this paper.

**Data availability**

Data will be made available on request.

**Acknowledgment**

D. Meister thanks the Stuttgart Center for Simulation Science (SimTech) for supporting him. The authors thank the anonymous reviewers for their comments which helped to improve this paper.

**Appendix A. Proof of Proposition 1**

First, note that the considered performance measure (2) is minimized if $\mathbb{E}[x(t)^\top L x(t)]$ is minimized for all $t \geq 0$. Second, the control input $u(t)$ can only utilize state information up to the latest triggering time instant, i.e., up to time $t_{\hat{k}(t)}$ with $\hat{k}(t) = \max\{k \in \mathbb{N}_0 \mid t_k \leq t\}$. We can write the corresponding global information set as $\mathcal{I}_t = \{x(s) \mid 0 \leq s \leq t_{\hat{k}(t)}\}$. Note that $\hat{k}(t)$ is time-dependent as it denotes the latest triggering instant with respect to time $t$. Thus, we know that there is no triggering instant between $t_{\hat{k}(t)}$ and $t$. Consequently, we can compute as follows

$$\begin{aligned}
\mathbb{E}[x(t)^\top L x(t)] &= \mathbb{E}\left[\mathbb{E}[x(t)^\top L x(t) | \mathcal{I}_t]\right] = \mathbb{E}\left[\mathbb{E}\left[\left(v(t) + \int_0^t u(s)\,\mathrm{d}s\right)^\top L(*) \bigg| \mathcal{I}_t\right]\right] \\
&= \mathbb{E}\left[(v(t) - v(t_{\hat{k}(t)}))^\top L(*)\right] + \sum_{k \in \mathbb{N}} H(t - t_k) \mathbb{E}\left[(v(t_k) - v(t_{k-1}))^\top L(*)\right] + 2\mathbb{E}\left[\mathbb{E}\left[(v(t) - v(t_{\hat{k}(t)}))^\top L \int_0^t u(s)\,\mathrm{d}s \bigg| \mathcal{I}_t\right]\right] \\
&\quad + 2\sum_{k \in \mathbb{N}} H(t - t_k) \mathbb{E}\left[(v(t_k) - v(t_{k-1}))^\top L \int_0^t u(s)\,\mathrm{d}s\right] + \mathbb{E}\left[\left(\int_0^t u(s)\,\mathrm{d}s\right)^\top L(*)\right] \\
&= \mathbb{E}\left[(v(t) - v(t_{\hat{k}(t)}))^\top L(*)\right] + \mathbb{E}\left[\left(v(t_{\hat{k}(t)}) + \int_0^t u(s)\,\mathrm{d}s\right)^\top L(*)\right],
\end{aligned} \tag{A.1}$$

where $(*)$ abbreviates the respective counterpart in the quadratic form, $v(t) = [v_1(t), \ldots, v_N(t)]^\top$, $H(\cdot)$ denotes the Heaviside step function, and $v(t_{\hat{k}(t)}) = \sum_{k \in \mathbb{N}} H(t - t_k)(v(t_k) - v(t_{k-1}))$. Moreover, we utilized the integrated version of the agent dynamics (1), the fact that the increments of a standard Brownian motion are independent, and that $u(s)$ must be independent of $v(\tilde{s})$ for $s, \tilde{s} \in (t_{\hat{k}(t)}, t]$. The latter yields $\mathbb{E}\left[(v(t) - v(t_{\hat{k}(t)}))^\top L \int_0^t u(s)\,\mathrm{d}s \bigg| \mathcal{I}_t\right] = 0$ for schemes with symmetric stopping times.

As both terms in (A.1) are non-negative, we arrive at the following optimality condition

$$\mathbb{E}\left[\left(v(t_{\hat{k}(t)}) + \int_0^t u(s)\,\mathrm{d}s\right)^\top L(*)\right] = \mathbb{E}\left[\left(x(t_{\hat{k}(t)}) + \int_{t_{\hat{k}(t)}}^t u(s)\,\mathrm{d}s\right)^\top L(*)\right] \overset{!}{=} 0 \quad \forall t \geq 0. \tag{A.2}$$

Evaluating the condition at $(t_k)_{k \in \mathbb{N}}$ yields $\mathbb{E}[x(t_k)^\top L x(t_k)] = 0$ which can be satisfied by ensuring

$$x(t_k)^\top L x(t_k) = 0 \quad \forall k \in \mathbb{N} \tag{A.3}$$

through the control inputs $u(t_k)$.

As explained in Section 2, the Laplacian matrix $L$ is positive semi-definite in the considered setup. Thus, we can decompose it according to $L = \tilde{L}^\top \tilde{L}$ which allows us to satisfy (A.2) by requiring $\tilde{L}\left(x(t_{\hat{k}(t)}) + \int_{t_{\hat{k}(t)}}^{t} u(s)\,\mathrm{d}s\right) = 0$ for all $t \geq 0$ or, equivalently,

$$L\left(x(t_{\hat{k}(t)}) + \int_{t_{\hat{k}(t)}}^{t} u(s)\,\mathrm{d}s\right) = 0 \quad \forall t \geq 0.$$

Evaluating at $(t_k)_{k \in \mathbb{N}}$ gives us $Lx(t_{\hat{k}(t)}) = 0$. Thus, for any $t \in \{t > 0 \mid \nexists k \in \mathbb{N} : t = t_k\}$, we have $L\int_{t_{\hat{k}(t)}}^{t} u(s)\,\mathrm{d}s = 0$, and, consequently,

$$Lu(t) = 0 \quad \forall t \in \{t > 0 \mid \nexists k \in \mathbb{N} : t = t_k\}. \tag{A.4}$$

We have now characterized a class of optimal causal control inputs with conditions (A.3) and (A.4). Note that the control input class stated in Proposition 1 fulfills these two conditions. In particular, (A.4) is satisfied by choosing $u(t) = 0$ for all $t \in \{t > 0 \mid \nexists k \in \mathbb{N} : t = t_k\}$. This appears as the practically most relevant case and provides some intuition on the controllers within the derived class. In addition, the control input class in Proposition 1 describes controllers that only utilize broadcast and local information $\mathcal{I}_{t,i}$ instead of global information $\mathcal{I}_t$ while still achieving the optimal performance level. Thus, we consider distributed controllers according to [27,28].

Note that all results shown in this paper generally hold for the class of control inputs satisfying (A.3) and (A.4). Moreover, a sufficient condition for optimality is all we need for our performance analysis since we are mainly interested in the best performance possible with any causal controller and not so much in the complete class of optimal causal controllers itself. □

**Appendix B. Proof of Lemma 1**

First, we compute as follows

$$\mathbb{E}\left[\int_0^M \frac{1}{2}\sum_{(i,j)\in\mathcal{E}} \left(x_i(t) - x_j(t)\right)^2 \mathrm{d}t\right] = \frac{1}{2}\sum_{(i,j)\in\mathcal{E}} \mathbb{E}\left[\int_0^M \left(x_i(t) - x_j(t)\right)^2 \mathrm{d}t\right]$$

$$= \frac{1}{2}\sum_{(i,j)\in\mathcal{E}} \left(\mathbb{E}\left[\sum_{k=1}^{m(M)} \int_{t_{k-1}}^{t_k} \left(x_i(t) - x_j(t)\right)^2 \mathrm{d}t\right] + \mathbb{E}\left[\int_{t_{m(M)}}^{M} \left(x_i(t) - x_j(t)\right)^2 \mathrm{d}t\right]\right), \tag{B.1}$$

where $(m(M))_{M \in [0,\infty)}$ is a renewal process for the renewal time sequence $(t_k)_{k \in \mathbb{N}}$. As the sequence of inter-event times is independent and identically distributed, the quantities

$$y_k^{(i,j)} := \int_{t_{k-1}}^{t_k} \left(x_i(t) - x_j(t)\right)^2 \mathrm{d}t$$

are independent and identically distributed as well. To see this, let $\bar{v}_i(t) = v_i(t) - v_i(t_k)$ for all $t \in [t_k, t_{k+1})$ and $i \in \{1,\ldots,N\}$. Note that $x_i(t) = x_i(t_k) + \bar{v}_i(t)$ for all $t \in [t_k, t_{k+1})$, $i \in \{1,\ldots,N\}$ and $x_i(t_k) = x_j(t_k)$ for all $i,j \in \{1,\ldots,N\}$.

By Wald's equation, we have $\mathbb{E}[\sum_{k=1}^{m(M)} y_k^{(i,j)}] = \mathbb{E}[m(M)]\,\mathbb{E}[y_1^{(i,j)}]$. In addition, the second term in (B.1) has the upper bound

$$\int_{t_{m(M)}}^{M} \left(x_i(t) - x_j(t)\right)^2 \mathrm{d}t \leq y_{m(M)+1}^{(i,j)}.$$

Dividing by $M$ and letting $M \to \infty$ yields

$$J = \frac{1}{2}\sum_{(i,j)\in\mathcal{E}} \lim_{M\to\infty} \frac{\mathbb{E}[m(M)]}{M} \cdot \mathbb{E}\left[y_1^{(i,j)}\right] = \frac{1}{\mathbb{E}[T]} \cdot \frac{1}{2}\sum_{(i,j)\in\mathcal{E}} \mathbb{E}\left[\int_0^T \left(x_i(t) - x_j(t)\right)^2 \mathrm{d}t\right],$$

since, by the renewal theorem and with $\mathbb{E}[T] < \infty$, $\lim_{M\to\infty} \frac{\mathbb{E}[m(M)]}{M} = \frac{1}{\mathbb{E}[T]}$. □

**Appendix C. Convergence of moments - Proof of Lemma 5**

We still have to show the following lemma to complete the proof of Lemma 5.

**Lemma 6.** *The limit theorem* (8) *also implies the convergence of the first and second moment, namely*

$$\mathbb{E}[X_N] \to \mathbb{E}[G] \quad \text{and} \quad \mathbb{E}[X_N^2] \to \mathbb{E}[G^2],$$

*where*

$$X_N := 2(\ln N)^2 \left(T_{\mathrm{ET}} - a_N\right)$$

*and $G$ is a Gumbel-distributed random variable. Furthermore, $a_N$ and $\kappa$ are defined as in* (8).

In the language of Lemma 6, the limit theorem (8) says that $X_N$ converges weakly to $G$.

**Proof.** The proof builds upon Lebesgue's dominated convergence theorem where we need to prove that $\mathbb{P}(X_N \geq r)$ and $2r\mathbb{P}(X_N \geq r)$ are upper bounded by integrable functions. We will show the existence of integrable majorants for $r \in \mathbb{R}$ by showing their existence on different ranges in $r$ whose union covers $\mathbb{R}$. Thereby, we arrive at a majorant for $r \in \mathbb{R}$ as the sum of the majorants for the considered subregions in $r$.

*Preliminaries*

As in the beginning of the proof for Lemma 5, we have

$$\mathbb{P}(T_j < w) = \mathbb{P}(\sup_{0 \leq t \leq w} |v_j(t)| \geq 1) = \mathbb{P}(\sup_{0 \leq t \leq 1} |v_j(t)| \geq w^{-1/2}).$$

Using

$$\mathbb{P}(\sup_{0 \leq t \leq 1} v_j(t) \geq w^{-1/2}) \leq \mathbb{P}(\sup_{0 \leq t \leq 1} |v_j(t)| \geq w^{-1/2}) \leq 2\mathbb{P}(\sup_{0 \leq t \leq 1} v_j(t) \geq w^{-1/2}),$$

as well as the fact that $\sup_{0 \leq t \leq 1} v_j(t)$ has the same distribution as $|v_j(1)|$, and the standard Gaussian tail estimate, we see that there exist $0 < \kappa_1 < \kappa_2 < \infty$ and $w_0 > 0$ such that, for all $0 < w < w_0$, we have

$$\frac{\kappa_1}{w^{-1/2}} e^{-w^{-1}/2} \leq \mathbb{P}(T_j < w) \leq \frac{\kappa_2}{w^{-1/2}} e^{-w^{-1}/2}. \tag{C.1}$$

Furthermore, we will use the inequalities

$$1 - x \leq (1 + x)^{-1} \leq 1 - x + x^2, \qquad x \geq 0, \tag{C.2}$$

and

$$\left(1 - \frac{\alpha}{N}\right)^N \leq e^{-\alpha}, \qquad \alpha \in \mathbb{R}. \tag{C.3}$$

*The range $0 \leq r \leq \frac{1}{2} \ln N$*

For $r > 0$, we have for large $N$ (uniformly in $r$)

$$w := \frac{r}{2(\ln N)^2} + a_N = \frac{r - \ln \frac{\kappa}{(2 \ln N)^{1/2}}}{2(\ln N)^2} + \frac{1}{2 \ln N} \geq \frac{1}{2 \ln N}, \tag{C.4}$$

i.e., $w^{-1/2} \leq (2 \ln N)^{1/2}$. Furthermore, since $r \leq \frac{1}{2} \ln N$, we have for large $N$ (uniformly in $r$)

$$(r - \ln \frac{\kappa}{(2 \ln N)^{1/2}})^2 \leq (r + \ln \ln N)^2 = r^2 + 2r \ln \ln N + (\ln \ln N)^2 \leq \frac{3}{4} r \ln N + (\ln \ln N)^2. \tag{C.5}$$

Therefore, the following holds

$$\mathbb{P}(X_N \geq r) = \mathbb{P}(T_1 \geq \frac{r}{2(\ln N)^2} + a_N)^N = \mathbb{P}(T_1 \geq w)^N \tag{C.6}$$

$$= \left(1 - \mathbb{P}(T_1 < w)\right)^N$$

$$\stackrel{(C.1)}{\leq} \left(1 - \frac{\kappa_1}{w^{-1/2}} \exp\left(-\frac{w^{-1}}{2}\right)\right)^N$$

$$\stackrel{(C.4)}{\leq} \left(1 - \frac{\kappa_1}{c_N} \exp\left(-\frac{1}{2}\left(\frac{r - \ln \frac{\kappa}{c_N}}{2(\ln N)^2} + \frac{1}{2 \ln N}\right)^{-1}\right)\right)^N = \left(1 - \frac{\kappa_1}{c_N} \exp\left(-(\ln N)\left(\frac{r - \ln \frac{\kappa}{c_N}}{\ln N} + 1\right)^{-1}\right)\right)^N$$

$$\stackrel{(C.2)}{\leq} \left(1 - \frac{\kappa_1}{c_N} \exp(-(\ln N)(1 - b_N + b_N^2))\right)^N = \left(1 - \frac{\kappa_1}{c_N} \frac{1}{N} \exp\left(r - \ln \frac{\kappa}{c_N} - \frac{(r - \ln \frac{\kappa}{c_N})^2}{\ln N}\right)\right)^N$$

$$= \left(1 - \frac{\kappa_1/\kappa}{N} \exp\left(r - \frac{(r - \ln \frac{\kappa}{c_N})^2}{\ln N}\right)\right)^N$$

$$\stackrel{(C.3)}{\leq} \exp\left(-\frac{\kappa_1}{\kappa} \exp\left(r - \frac{(r - \ln \frac{\kappa}{c_N})^2}{\ln N}\right)\right)$$

$$\stackrel{(C.5)}{\leq} \exp\left(-\frac{\kappa_1}{\kappa} \exp\left(\frac{r}{4} - 1\right)\right), \tag{C.7}$$

where $b_N = \frac{r - \ln(\kappa/c_N)}{\ln N}$ and $c_N = (2 \ln N)^{1/2}$. In addition, we utilized $r \in [0, \frac{1}{2} \ln N]$ for applying (C.1), (C.2), (C.4) and (C.5).

*The range $r > (\ln N)^2$*

On this range, we have that for $N$ large enough (uniformly in $r$)

$$w = \frac{r - \ln \frac{\kappa}{(2\ln N)^{1/2}}}{2(\ln N)^2} + \frac{1}{2\ln N} \geq \frac{r}{2(\ln N)^2}, \tag{C.8}$$

and $r > (\ln N)^2$ implies $w \geq 1/2$. Here, we can use the standard small deviation estimate as in [36, (1.3)]: For a constant $c > 0$ and all $w \geq 1/2$, we have

$$\mathbb{P}(T_1 \geq w) = \mathbb{P}(\sup_{0 \leq t \leq w} |v_1(t)| \leq 1) \leq e^{-cw}. \tag{C.9}$$

Therefore, continuing in (C.6), we obtain

$$\mathbb{P}(X_N \geq r) = \mathbb{P}(T_1 \geq w)^N \stackrel{(C.9)}{\leq} e^{-cwN} \stackrel{(C.8)}{\leq} \exp\left(-c\frac{rN}{2(\ln N)^2}\right) \leq e^{-r}, \tag{C.10}$$

where we utilized $r > (\ln N)^2$ to apply (C.9).

*The range $\frac{1}{2}\ln N \leq r \leq (\ln N)^2$*

Here, for $N$ large enough (uniformly in $r$), we have

$$w = \frac{r - \ln \frac{\kappa}{(2\ln N)^{1/2}}}{2(\ln N)^2} + \frac{1}{2\ln N} \geq \frac{3}{4\ln N}. \tag{C.11}$$

Therefore, we obtain

$$\mathbb{P}(X_N \geq r) = \mathbb{P}(T_1 \geq w)^N \stackrel{(C.11)}{\leq} \mathbb{P}\left(T_1 \geq \frac{3}{4\ln N}\right)^N = \left(1 - \mathbb{P}\left(T_1 < \frac{3}{4\ln N}\right)\right)^N$$

$$\stackrel{(C.1)}{\leq} \left(1 - \frac{\kappa_1}{(\frac{4}{3}\ln N)^{1/2}} e^{-\frac{2}{3}\ln N}\right)^N = \left(1 - \frac{\kappa_1}{N^{2/3}(\frac{4}{3}\ln N)^{1/2}}\right)^N = \left(1 - \frac{\kappa_1 N^{1/3}}{N(\frac{4}{3}\ln N)^{1/2}}\right)^N$$

$$\stackrel{(C.3)}{\leq} \exp\left(-\frac{\kappa_1 N^{1/3}}{(\frac{4}{3}\ln N)^{1/2}}\right) \leq \exp(-N^{1/6}) \leq \exp\left(-e^{\sqrt{r}/6}\right), \tag{C.12}$$

where we utilized $\frac{1}{2}\ln N \leq r$ to apply (C.11) and $r \leq (\ln N)^2$ for the last step.

Putting the three estimates (C.7), (C.10) and (C.12) together shows that we can find an integrable majorant for $r \mapsto \mathbb{P}(X_N \geq r)$. Therefore, using the limit theorem (8)

$$\mathbb{E}\left[X_N \mathbb{1}_{X_N \geq 0}\right] = \int_0^\infty \mathbb{P}(X_N \geq r)\,dr \to \int_0^\infty \mathbb{P}(G \geq r)\,dr = \mathbb{E}\left[G\mathbb{1}_{G>0}\right]. \tag{C.13}$$

Similarly, the three estimates above show that an integrable majorant for $r \mapsto 2r\mathbb{P}(X_N \geq r)$ can be found giving

$$\mathbb{E}\left[X_N^2 \mathbb{1}_{X_N \geq 0}\right] = \int_0^\infty 2r\mathbb{P}(X_N \geq r)\,dr \to \int_0^\infty 2r\mathbb{P}(G \geq r)\,dr = \mathbb{E}\left[G^2 \mathbb{1}_{G>0}\right]. \tag{C.14}$$

*The range $r < 0$*

We finally handle $\mathbb{E}\left[X_N \mathbb{1}_{X_N < 0}\right]$. Note that

$$-\mathbb{E}\left[X_N \mathbb{1}_{X_N < 0}\right] = \int_0^\infty \mathbb{P}(-X_N > r)\,dr = \int_{-\infty}^0 \mathbb{P}(X_N < r)\,dr,$$

and since $X_N \geq -\ln N + \ln(\kappa/(2\ln N)^{1/2}) =: r_{\min}$, the integral is actually on the range $[r_{\min}, 0]$. This time, we will find an integrable majorant for $r \mapsto \mathbb{P}(X_N < r)$ on this range.

Recall from (C.6) that

$$\mathbb{P}(X_N < r) = 1 - \mathbb{P}(X_N \geq r) = 1 - \mathbb{P}(T_1 \geq w)^N = 1 - (1 - \mathbb{P}(T_1 < w))^N,$$

where, as above,

$$w = w_r = \frac{r - \ln \frac{\kappa}{(2\ln N)^{1/2}}}{2(\ln N)^2} + \frac{1}{2\ln N} \in \left[0, \frac{1}{2\ln N} - \frac{\ln \frac{\kappa}{(2\ln N)^{1/2}}}{2(\ln N)^2}\right]. \tag{C.15}$$

Thus, it suffices to find an integrable majorant for $r \mapsto 1 - (1 - \mathbb{P}(T_1 < w_r))^N$.

Recall that for $x \in [0, 1]$, we have $1 - (1 - x)^N \leq Nx$. Therefore, we obtain

$$1 - (1 - \mathbb{P}(T_1 \leq w))^N \leq N\mathbb{P}(T_1 < w)$$

$$\stackrel{(C.1)}{\underset{(C.15)}{\leq}} N \frac{\kappa_2}{w^{-1/2}} \exp\left(-\frac{w^{-1}}{2}\right) = N \frac{\kappa_2}{w^{-1/2}} \exp\left(-\frac{1}{2}\left(\frac{r - \ln\frac{\kappa}{c_N}}{2(\ln N)^2} + \frac{1}{2\ln N}\right)^{-1}\right)$$

$$= N \frac{\kappa_2}{w^{-1/2}} \exp\left(-\ln N \left(\frac{r - \ln\frac{\kappa}{c_N}}{\ln N} + 1\right)^{-1}\right)$$

$$\stackrel{(C.2)}{\leq} N \frac{\kappa_2}{w^{-1/2}} \exp\left(-\ln N \left(1 - \frac{r - \ln\frac{\kappa}{c_N}}{\ln N}\right)\right) = \frac{\kappa_2}{w^{-1/2}} \exp\left(r - \ln\frac{\kappa}{c_N}\right)$$

$$= \frac{\kappa_2}{w^{-1/2}} e^r \frac{c_N}{\kappa} = e^r \frac{\kappa_2}{\kappa} (w \cdot 2\ln N)^{1/2}$$

$$\stackrel{(C.15)}{\leq} e^r \frac{\kappa_2}{\kappa} \sqrt{2},$$

for large $N$ (uniformly in $r$) and $c_N = (2\ln N)^{1/2}$.

This gives an integrable majorant for $r \mapsto \mathbb{P}(X_N < r)$ for $r \in (-\infty, 0]$. Therefore, we have

$$-\mathbb{E}\left[X_N \mathbb{1}_{X_N < 0}\right] = \int_{-\infty}^0 \mathbb{P}(X_N < r)\,dr \to \int_{-\infty}^0 \mathbb{P}(G < r)\,dr = -\mathbb{E}\left[G\mathbb{1}_{G<0}\right].$$

The same argument applies to $\mathbb{E}\left[X_N^2 \mathbb{1}_{X_N<0}\right]$ since the above estimate can also be used for $r \mapsto 2(-r)\mathbb{P}(X_N < r)$.

This together with (C.13) and (C.14) shows that $\mathbb{E}[X_N] \to \mathbb{E}[G]$ and $\mathbb{E}[X_N^2] \to \mathbb{E}[G^2]$, as required. □

**Appendix D. Estimation of the cost ratio in simulation**

Let us provide details on the utilized approach to estimate the event-triggered cost in the cost ratio and, subsequently, on the methodology to obtain the confidence interval depicted in Fig. 3.

Consider the Bessel process $R(t) := \sqrt{\sum_{i=1}^N v_i(t)^2}$ of dimension $N$ started at $R(0) = 0$.

**Lemma 7.** *For any stopping time $T$ satisfying the assumptions in Lemmas 1 and 2, we have*

$$Q(T) = \frac{|\mathcal{E}|}{2N(N+2)} \mathbb{E}\left[R(T)^4\right].$$

**Proof.** The stochastic differential equation solved by $(R(t))_{t\in[0,\infty)}$ is given by

$$dR(t) = \frac{N-1}{2R(t)}dt + dv(t),$$

where $v(t)$ is a standard Brownian motion. Let $Af(x) := \frac{N-1}{2x}f'(x) + \frac{1}{2}f''(x)$ be the infinitesimal generator of the Markov process $(R(t))_{t\in[0,\infty)}$. Then, Dynkin's formula says that, for any stopping time $T$,

$$\mathbb{E}[f(R(T))] = \mathbb{E}\left[\int_0^T Af(R(t))\,dt\right].$$

Let $f(x) = x^4$. Then, $Af(x) = \frac{N-1}{2x}4x^3 + \frac{4\cdot 3}{2}x^2 = 2(N+2)x^2$. This yields

$$\mathbb{E}\left[R(T)^4\right] = 2(N+2)\mathbb{E}\left[\int_0^T R(t)^2\,dt\right].$$

Utilizing

$$\mathbb{E}\left[\int_0^T R(t)^2\,dt\right] = N\mathbb{E}\left[\int_0^T v_1(t)^2\,dt\right] = \frac{N}{|\mathcal{E}|}Q(T),$$

based on Lemma 2 gives the stated formula for $Q(T)$. □

We can leverage this lemma to estimate $Q_{\text{ET}}(\Delta)$ in simulation and thereby arrive at an estimate for $J_{\text{ET}}(\Delta)$. This is required to obtain the cost ratio results in Section 5. In particular, note that this lemma allows for the explicit cancellation of $|\mathcal{E}|$ from the cost ratio. Thus, the obtained simulation results hold for any graph topology in the given framework.

At last, let us present the methodology to obtain the confidence interval in Fig. 3.

**Lemma 8.** *Let the estimates for $\mathbb{E}\left[R(T_{\mathrm{ET}})^4\right]$ and $\mathbb{E}\left[T_{\mathrm{ET}}\right]$ be*

$$g_1 = \frac{1}{n_1} \sum_{i=1}^{n_1} R(T_{\mathrm{ET}})_i^4, \quad g_2 = \frac{1}{n_2} \sum_{i=1}^{n_2} T_{\mathrm{ET},i},$$

*respectively, where $n_1, n_2 \in \mathbb{N}$ denote the number of samples $R(T_{\mathrm{ET}})_i^4$ and $T_{\mathrm{ET},i}$ obtained from independent Monte Carlo simulations. Moreover, let $[g_1^{\mathrm{L}}, g_1^{\mathrm{R}}]$, $[g_2^{\mathrm{L}}, g_2^{\mathrm{R}}]$ be the corresponding confidence intervals with confidence level $\gamma$. Then, we can estimate the confidence interval $[\alpha^{\mathrm{L}}, \alpha^{\mathrm{R}}]$ with*

$$\mathbb{P}\left(\alpha^{\mathrm{L}} \leq \frac{J_{\mathrm{ET}}(\Delta)}{J_{\mathrm{TT}}(\mathbb{E}[T_{\mathrm{ET}}(\Delta)])} \leq \alpha^{\mathrm{R}}\right) \geq \gamma^2$$

*according to*

$$\alpha^{\mathrm{L}} = \frac{g_1^{\mathrm{L}}}{N(N+2)(g_2^{\mathrm{R}})^2}, \quad \alpha^{\mathrm{R}} = \frac{g_1^{\mathrm{R}}}{N(N+2)(g_2^{\mathrm{L}})^2}.$$

**Proof.** With Lemma 7, we can estimate the cost ratio as

$$\frac{J_{\mathrm{ET}}(\Delta)}{J_{\mathrm{TT}}(\mathbb{E}[T_{\mathrm{ET}}(\Delta)])} \approx \frac{g_1}{N(N+2)g_2^2}.$$

Given the confidence intervals $[g_1^{\mathrm{L}}, g_1^{\mathrm{R}}]$, $[g_2^{\mathrm{L}}, g_2^{\mathrm{R}}]$, we find

$$\mathbb{P}\left(\alpha^{\mathrm{L}} \leq \frac{J_{\mathrm{ET}}(\Delta)}{J_{\mathrm{TT}}(\mathbb{E}[T_{\mathrm{ET}}(\Delta)])} \leq \alpha^{\mathrm{R}}\right) \geq \mathbb{P}(g_1 \in [g_1^{\mathrm{L}}, g_1^{\mathrm{R}}] \wedge g_2 \in [g_2^{\mathrm{L}}, g_2^{\mathrm{R}}])$$
$$= \mathbb{P}(g_1 \in [g_1^{\mathrm{L}}, g_1^{\mathrm{R}}]) \cdot \mathbb{P}(g_2 \in [g_2^{\mathrm{L}}, g_2^{\mathrm{R}}]) = \gamma^2,$$

which finishes the proof. □

Since we have large numbers of samples $n_1, n_2$, we can assume the samples $R(T_{\mathrm{ET}})_i^4, T_{\mathrm{ET},i}$ to be normally distributed. This allows us to apply standard statistical methods to determine the confidence intervals $[g_1^{\mathrm{L}}, g_1^{\mathrm{R}}]$, $[g_2^{\mathrm{L}}, g_2^{\mathrm{R}}]$, e.g., for a confidence level $\gamma = 0.975$. Consequently, the confidence interval $[\alpha^{\mathrm{L}}, \alpha^{\mathrm{R}}]$ according to Lemma 8 corresponds to a confidence level of at least $\gamma^2 = 0.975^2 \approx 0.95$.